\documentclass[traditabstract]{aa} 

\newcommand{\unit}[1]{\ensuremath{\mathrm{\thinspace #1}}}
\newcommand{\od}[2]{\frac{\mathrm{d}#1}{\mathrm{d}#2}}
\newcommand{\reffig}[1]{\figurename~\ref{#1}}
\newcommand{\MMR}[2]{\mbox{#1:#2}}

\def\mearth{\mathrm{\thinspace M_\oplus}}

\usepackage{graphicx}
\usepackage{txfonts}

\usepackage{natbib}
\usepackage{aas_macros}

\begin{document}

\titlerunning{Shifted CZs}
\authorrunning{Cossou et al.}

\title{Convergence zones for Type I migration: an inward shift for multiple planet systems}

\author{Christophe Cossou\inst{1,2}, Sean N. Raymond\inst{1,2}, \and Arnaud Pierens\inst{1,2}}

\institute{
Univ. Bordeaux, LAB, UMR 5804, F-33270, Floirac, France.
\and
CNRS, LAB, UMR 5804, F-33270, Floirac, France
}

\date{5 December 2012 / 7 February 2013}

\abstract{

Earth-mass planets embedded in gaseous protoplanetary disks undergo Type I orbital migration.  In radiative disks an additional component of the corotation torque scaling with the entropy gradient across the horseshoe region can counteract the general inward migration, Type I migration can then be directed either inward or outward depending on the local disk properties.  Thus, special locations exist in the disk toward which planets migrate in a convergent way.  Here we present N-body simulations of the convergent migration of systems of low-mass (M=1-10$\mearth$) planets.  We show that planets do not actually converge in convergence zones.  Rather, they become trapped in chains of mean motion resonances (MMRs).  This causes the planets' eccentricities to increase to high enough values to affect the structure of the horseshoe region and weaken the positive corotation torque.  The zero-torque equilibrium point of the resonant chain of planets is determined by the sum of the attenuated corotation torques and unattenuated differential Lindblad torques acting on each planet.  The effective convergence zone is shifted inward.  Systems with several planets can experience stochastic migration as a whole due to continuous perturbations from planets entering and leaving resonances.  

}{}{}{}{}


\maketitle

\section{Introduction}

An Earth- to Neptune-mass planet embedded in a gaseous protoplanetary disk excites density waves in the disk \citep{goldreich1979excitation}, and the back-reaction of these perturbations on the planet's orbit causes it to undergo so-called Type I migration \citep{ward1997survival}.  In isothermal disks Type I migration is driven only by the differential Lindblad torque and is inward-directed and fast \citep{tanaka2002three}.  In radiative disks, a torque arising from material in a planet's horseshoe region can counteract the differential Lindblad torque such that migration can be directed either inward or outward (\cite{paardekooper2006halting}; \cite{kley2008migration}).  There are locations within disks for which the migration is convergent, and we refer to these as convergence zones \citep[CZs;][]{lyra2010orbital, paardekooper2011torque}.  

At a convergence zone, the positive corotation torque is in exact balance with the negative differential Lindblad torque, such that the total torque experienced by the planet is zero. It has been proposed that CZs may act to concentrate planetary embryos and build large cores \citep{lyra2010orbital, horn2012orbital}.  As planets migrate toward the CZ they become trapped in mean-motion resonances (MMRs), which prevent unlimited accretion (\cite{morbidelli2008building}, \cite{sandor2011formation}).  However, collisions do occur once embryos are trapped in a long enough resonant chain to destabilize it.  In addition, turbulence may act to break resonances and enhance accretion.  

\cite{bitsch2010orbital} show that the corotation torque is attenuated when a planet has an eccentricity such that its orbit crosses a significant fraction of the width of the horseshoe region.  When two planets become trapped in resonance due to convergent migration they excite and sustain non-zero eccentricities despite eccentricity damping by the disk \citep[e.g.][]{cresswell2008three}. This in turn should decrease the magnitude of the corotation torque and thus modify the balance between the differential Lindblad and corotation torques and the subsequent migration.  

Here we present simulations of the convergent migration of low-mass (M=1-10$\mearth$) planets in idealized gas disks.  We use a simple model for the influence of the eccentricity on the corotation torque.  We show that planets do not actually converge at the convergence zone but  are instead systematically driven inward to a shifted equilibrium position with zero net torque.  The location of the equilibrium position depends on the eccentricity excitation of the planets due to mutual perturbations.  

The paper is laid out as follows.  In \S 2 we present our numerical methods and assumptions.  In \S 3 we present results for two planets.  In \S 4 we present results for 3-10 planets.  We discuss our results in \S 5.  

\section{Methods}
\begin{figure}[htb]
\centering
\includegraphics[width=0.49\linewidth]{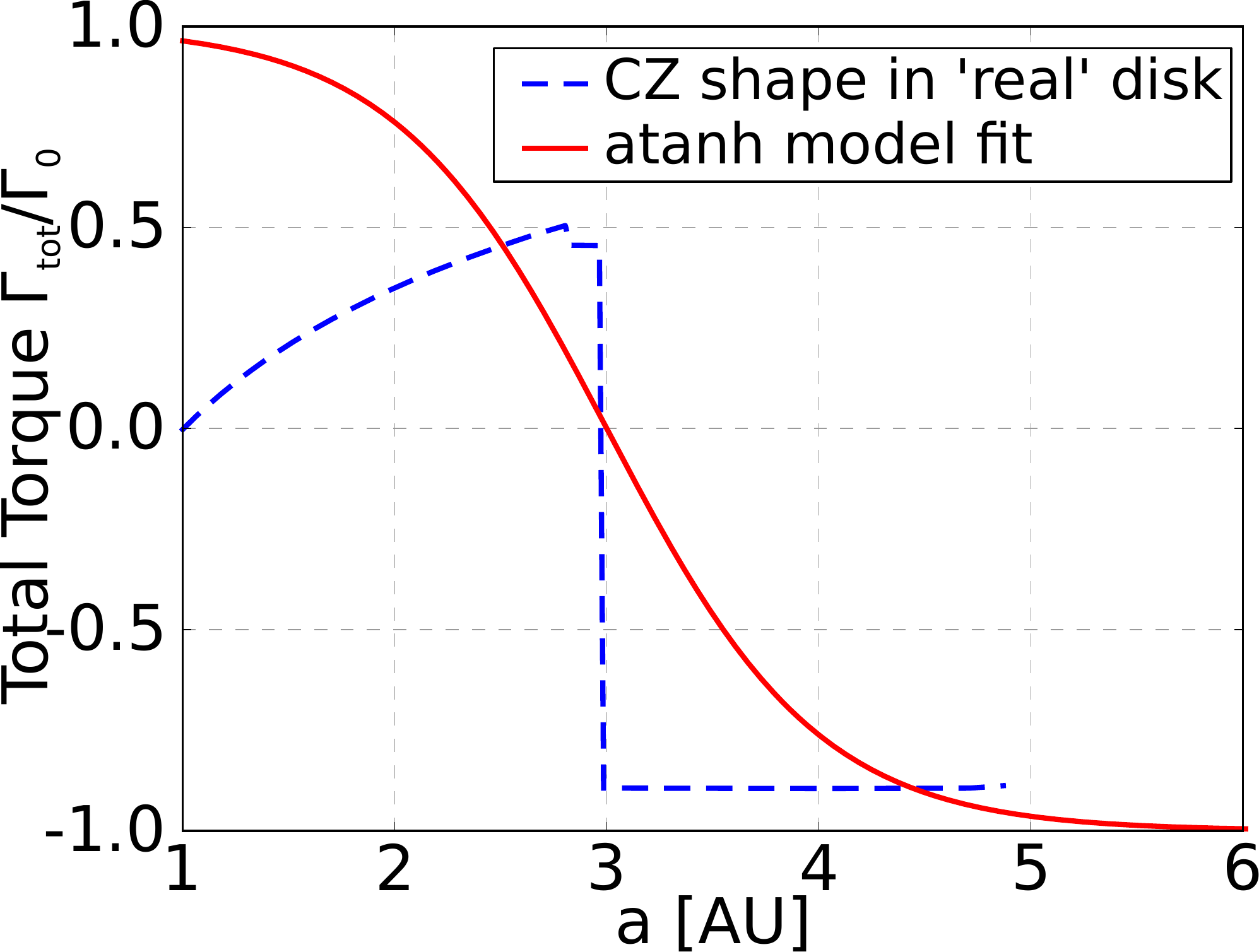}\hfill
\includegraphics[width=0.49\linewidth]{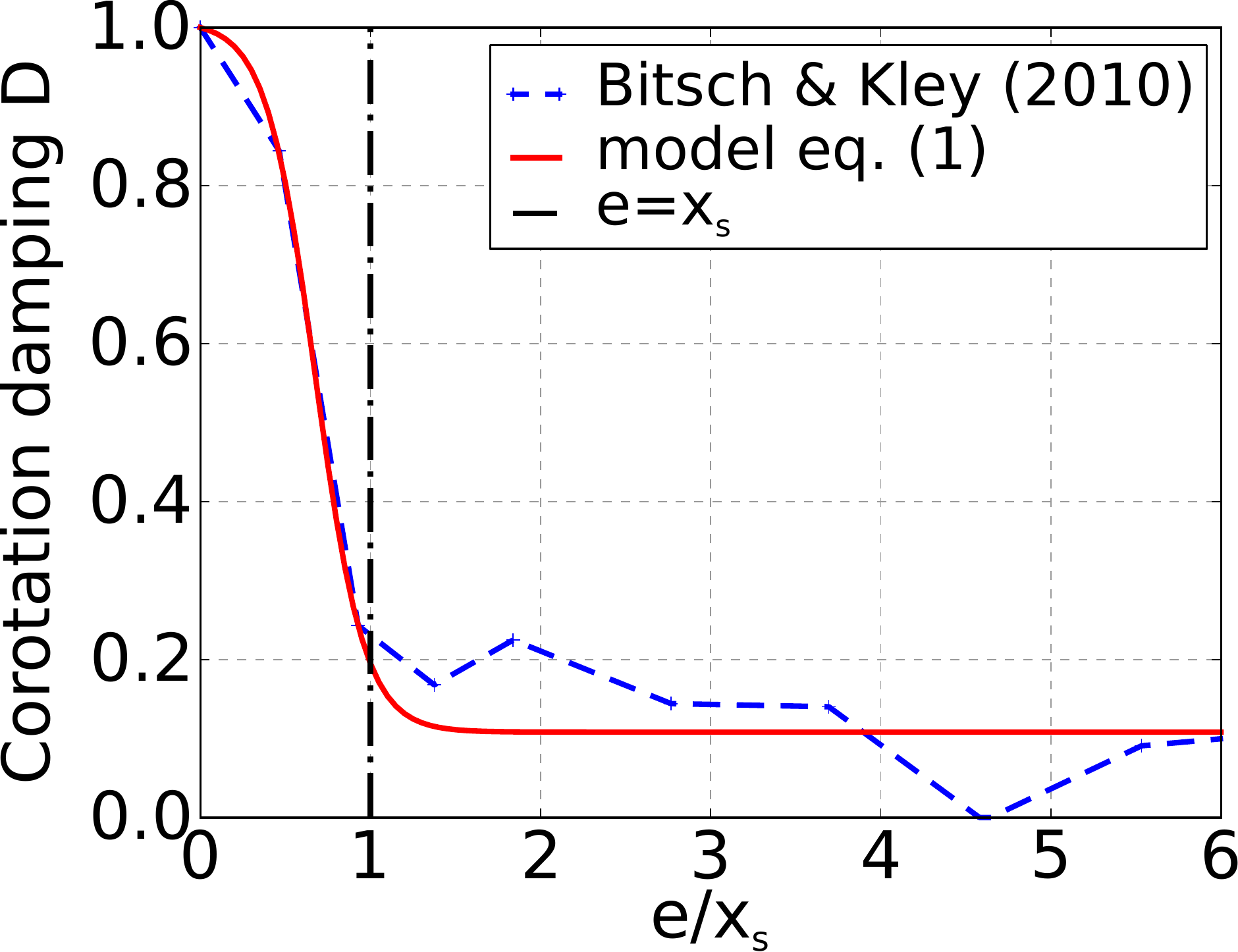}
\caption{{\bf Left:} The total torque profile of our standard disk in units of $\Gamma_0$ is shown in red. The dashed blue curve shows the true torque profile for a $10m_\oplus$ planet near an opacity transition, calculated using the equations in~\cite{paardekooper2011torque}. {\bf Right:} Decrease in the corotation torque $\Gamma_C$ as a function of planetary eccentricity $e$. We assume that the damping ($0<D<1$) of the corotation torque as a function of eccentricity is the same for isothermal and fully radiative disc. Then, we derived $D$ from Fig. 2 of \cite{bitsch2010orbital} by taking the difference between their fully radiative and isothermal case, and normalizing to 1 the ``$e=0$ case".}

\label{fig:eccentricity-influence}
\end{figure}

We use an artificial convergence zone (CZ) that mimics a mass-independent CZ (whose position does not depend on the mass of the planet) at an opacity transition, see (left panel of) \reffig{fig:eccentricity-influence}, where there is an abrupt change from outward to inward migration~\citep[see, e.g., ][]{masset2011type}. A step function was not used because the step in the `real' profile is only due to the opacity table not being smoothed. The location of the CZ was 3 AU. Inside 3 AU the total torque is positive and equal to $\Gamma_0 = \left(\frac{q}{h}\right)^2\Sigma_p {r_p}^4 {\Omega_p}^2$. Outside 3 AU the total torque is $-\Gamma_0$. Here $q$ is the ratio between the mass of the planet and the star, $h$ is the aspect ratio that depend on the temperature profile but is typically $0.05$, $\Sigma_p$, $r_p$, and $\Omega_p$ are respectively the surface density, the orbital distance, and the angular speed for the planet. The total torque is a sum of the differential Lindblad torque $\Gamma_L$ --- assumed to remain constant --- and the corotation torque $\Gamma_C$. The main interest of the artificial CZ is to get rid of the very complex shape of the real profile, to only retain the CZ, and nothing more.

\cite{bitsch2010orbital} show that as the eccentricity of a body increases, the structure of its horseshoe region is modified and $\Gamma_C$ decreases.  We implemented a simple formula that reproduces the general influence of the eccentricity on $\Gamma_C$ by a simple fit to the 3D simulations of \citep{bitsch2010orbital}:
\begin{equation}
D = \frac{\Gamma_C(e)}{\Gamma_C (e=0)} = 1 + a \cdot \left[\tanh(c) - \tanh\left(\frac{b * e}{x_s}+c\right)\right],\label{eq:eccentricity-influence}
\end{equation}
where $x_s$ represents the half-width of the horseshoe region divided by the semi major axis, $e$ is the planet's eccentricity, and our fit produced $a = 0.45$, $b = 3.46$, and $c = -2.34$. We define $x_s$ as \citep[see][eq. 44]{paardekooper2010torque}:
\begin{equation}
x_s = \frac{1.1}{\gamma^{1/4}} \sqrt{\frac{q}{h}}
\end{equation}
where $\gamma$ is the adiabatic index, $q$ the planet-to-stellar mass ratio, and $h$  the disk aspect ratio. The right hand panel of \reffig{fig:eccentricity-influence} shows that our simple equation is a good match to hydrodynamical simulations of the effect of $e$ on $\Gamma_C$, especially at small $e$, although the simulation data points are sparse and appear to include some random fluctuations.

To perform our simulations we used a modified version of the Mercury integrator \citep{chambers1999hybrid}.  We included both the artificial convergence zone (where $\Gamma = \Gamma_L+\Gamma_C$) and $\Gamma_C$ varies with the planet's eccentricity following Eq.\ref{eq:eccentricity-influence}.  Eccentricity and inclination damping is done using formulas from \cite{cresswell2008three}.  We assumed the presence of a gas disk with surface density profile $\Sigma = 500 \left(r/1\unit{AU}\right)^{-1/2} \unit{g.cm^{-2}}$ (used to calculate $\Gamma_0$ and the damping).  

To implement the migration caused by the disk's torque $\Gamma$, we note that $\Gamma=\od{J}{t}$, and we define an extra acceleration $a_m$ defined as $a_m = - \frac{v}{t_m}$ \citep[eq. (14)]{cresswell2008three} where $v$ is the planet's velocity, and $t_m=J/\od{J}{t}$ the migration time ($J$ is the angular momentum).

In all simulations planets started on low eccentricity and low inclination orbits.  We ran each simulation for three million years with a time step between 0.4 and 3 days.

\section{Two-planet case}

\begin{figure}[htb]
\centering
\includegraphics[width=\linewidth]{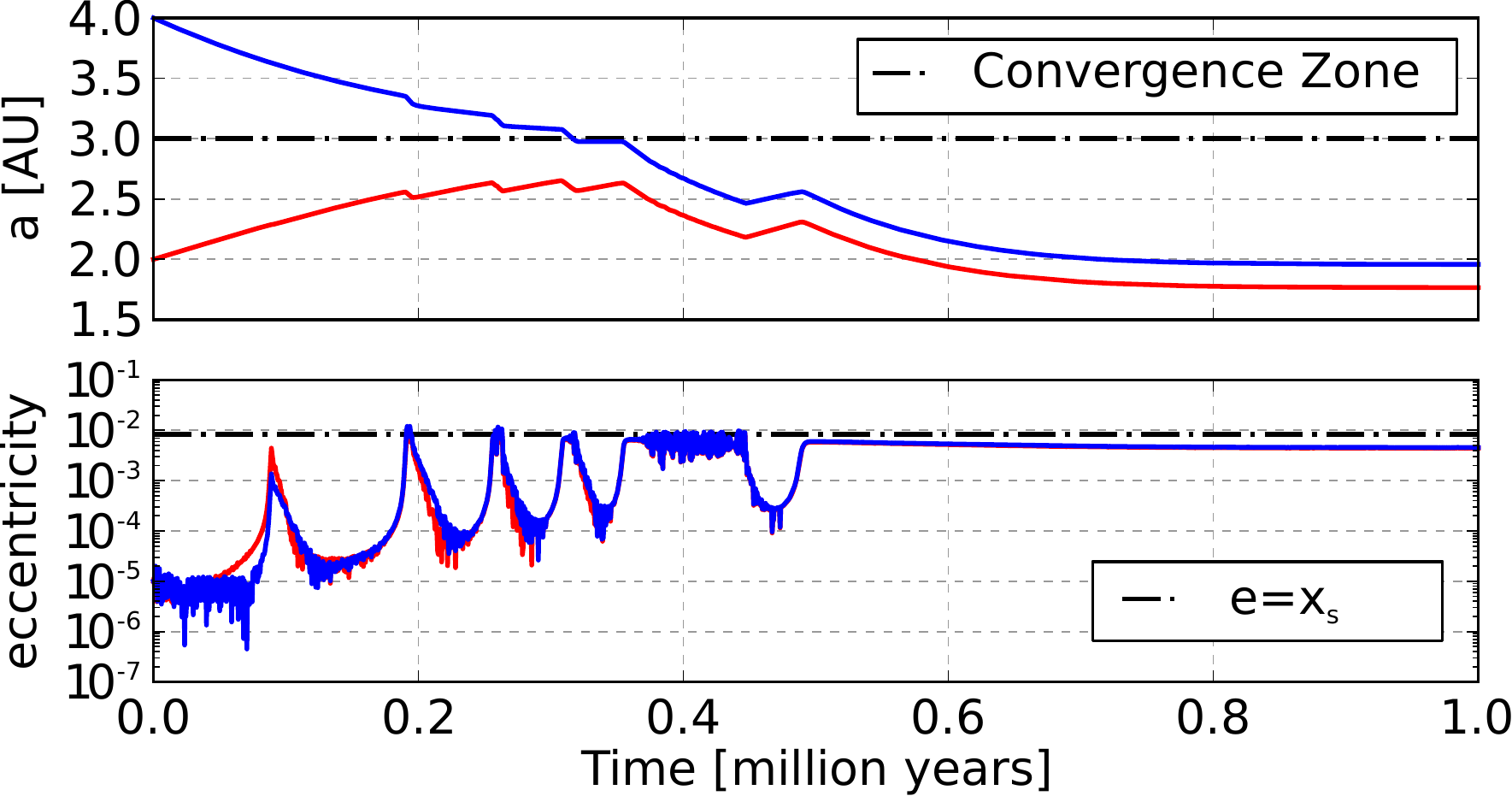}
\caption{Simulation of the convergent migration of two $1\unit{M_\oplus}$ planets toward the CZ at 3 AU including the $\Gamma_C-e$ feedback illustrated in the right panel of \reffig{fig:eccentricity-influence}.}
\label{fig:two-planets}
\end{figure}

Figure~\ref{fig:two-planets} shows the evolution of two $1\unit{M_\oplus}$ planets initially on opposite sides of the CZ at 3 AU.  As they approach each other, the two planets cross a series of MMRs and are eventually trapped in the \MMR{7}{6} resonance.  The eccentricities of the two planets reach an equilibrium between resonant excitation and damping by the disk.  This equilibrium eccentricity is 0.5 times the horseshoe semi-width $x_s$ and damps the corotation torque to roughly 80\% of its zero-eccentricity value. 

The planets reach a stable orbital configuration at 1.77 and 1.96 AU, with both planets interior to the pre-imposed CZ. Given their eccentricities, the innermost planet CZ is shifted to 1.95 AU while it is 1.74 for the outer (1.96 AU) planet. In this context, the shift comes from  the balance between the unchanged negative Lindblad torque and the attenuated positive corotation torque.  \emph{The two-planet system stabilizes at a zero net torque point --- even though each planet feels a nonzero torque and neither reaches its single-planet CZ --- that is shifted inward from the nominal CZ by $>1\unit{AU}$. }

It is clear that the planets' eccentricities --- excited by planet-planet interactions --- are the key factor in determining the strength of the corotation torque and the location of the effective convergence zone.  For two equal-mass planets the same qualitative behavior occurs regardless of the planet mass, although the evolution for higher mass planets is faster.  In addition, the eccentricity excitation varies both with the planet mass and which resonance the planets occupy: higher eccentricities imply stronger damping of $\Gamma_C$ and a closer-in effective CZ.

\subsection{Effect of mass ratio}
\begin{figure}[htb]
\centering
\includegraphics[width=0.95\linewidth]{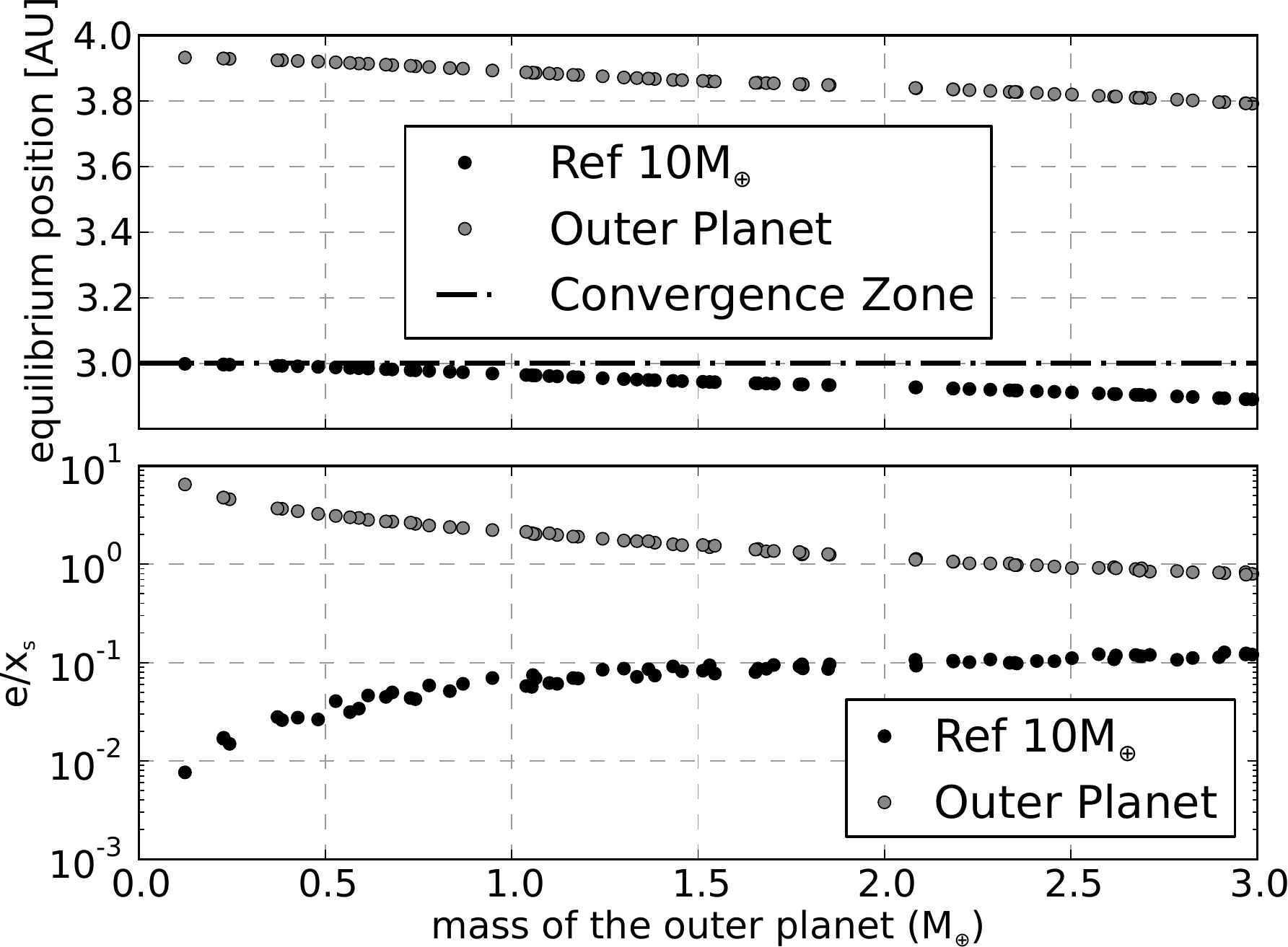}
\caption{Outcome of a set of simulations that started with a $10 \mearth$ planet at 3 AU and a second planet of $0.1-3 \mearth$ at 4 AU.  The panels show the equilibrium position of the planets (top) and the eccentricity normalized to the horseshoe semi-width $e/x_s$ (bottom) as a function of the outer planet's mass.}\label{fig:mass_ratio_final_pos}
\end{figure}

We now study the case of two unequal-mass planets. Figure~\ref{fig:mass_ratio_final_pos} shows the outcome of a simple experiment in which a $10\unit{M_\oplus}$ planet was placed at the CZ at $3\unit{AU}$ and a second planet was placed at $4\unit{AU}$.  The mass of the second planet was varied in a set of simulations from $0.1$ to $10\unit{M_\oplus}$.  

In \reffig{fig:mass_ratio_final_pos} the outer planet remained in \MMR{3}{2} resonance.  In this situation, the final location of the planets is determined by their relative masses or, for this experiment, the outer planet's mass.  The more massive the outer planet, the closer in the two planets' effective CZ.  The inner planet's eccentricity increases for a more massive outer planet, causing a corresponding decrease in $\Gamma_C$ and a more drastic inward shift.  Given that each planet has a different mass and eccentricity, they have different effective CZ (one per $e/x_s$ value).  The magnitude of the overall inward shift, however, is mainly determined by the most massive planet.  

Figure~\ref{fig:mass_ratio_final_pos} only shows a subset of the simulations run in this numerical experiment.  For higher outer planet masses, the two planets were trapped in different MMRs, resulting in discontinuities in the equilibrium positions and mean eccentricity values.  However, the behavior of the two planet system remained qualitatively similar.  

\subsection{Resonance effect}
The order of MMR is also important (for an MMR of $(p+q)/p$, $p$ is the order). Two planets in \MMR{3}{2} resonance will have higher eccentricities than the same planets in \MMR{11}{10} resonance . The simple explanation is that lower order resonances have more frequent conjunctions, hence stronger resonant perturbations (see \cite{murray2000solar} for more details).  The MMR in which a two-planet system is captured depends on the relative migration speed and the rate of eccentricity damping (e.g., \cite{mustill2011general}), both determined by the disk and torque profiles, and the initial positions of the planets.  

\begin{figure}[htb]
\centering
\includegraphics[width=0.95\linewidth]{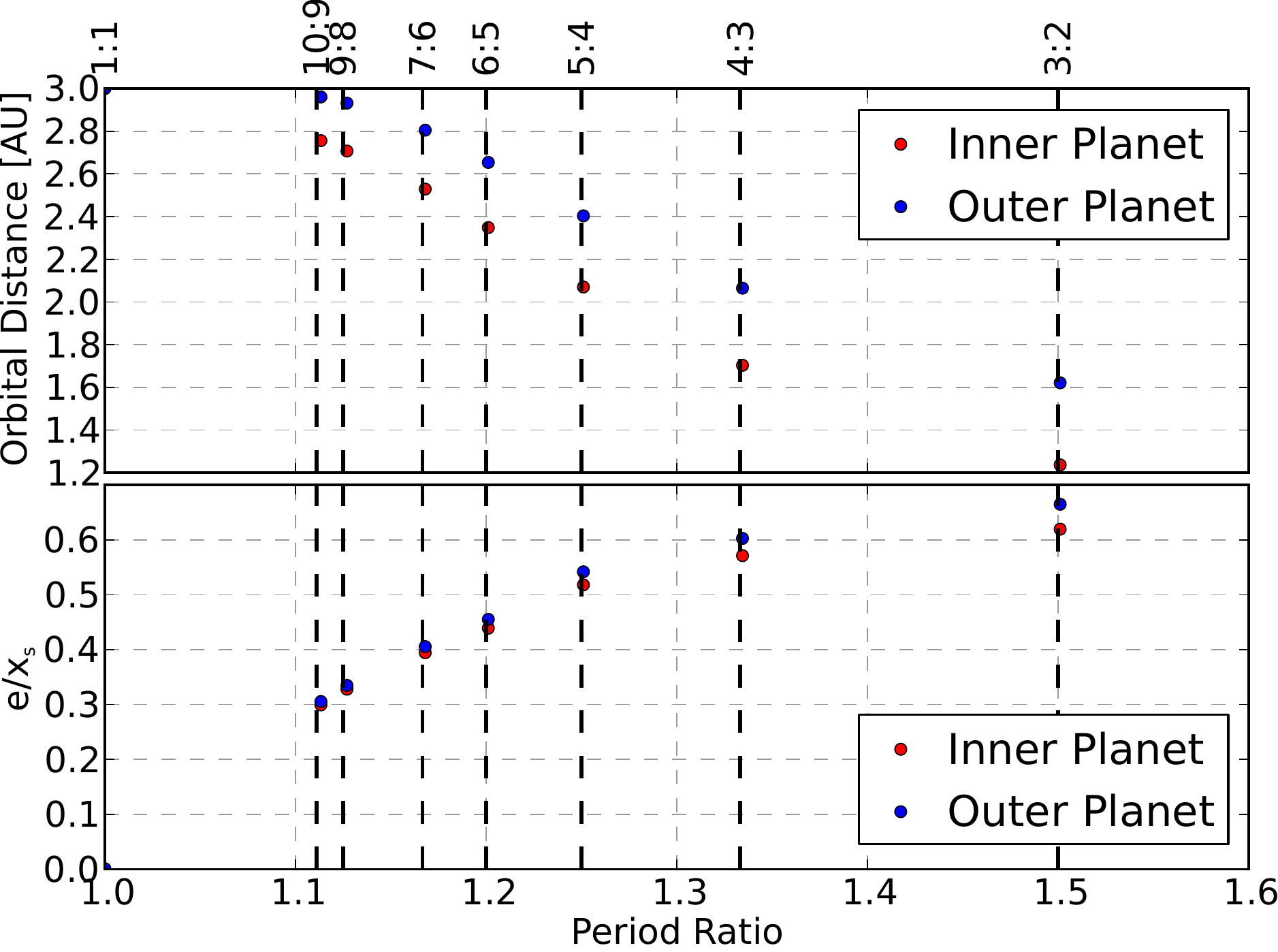}
\caption{Final semimajor axes (top) and eccentricities (bottom) of two $3 \mearth$ planets trapped in different MMRs. For a $3\mearth$ planet, the semiwidth of the horseshoe region is about $0.014$.}\label{fig:influence_of_MMR}
\end{figure}

We performed a set of 100 simulations (1 Myr each) with two $3 \mearth$ planets that were each initially placed randomly between 1 and 10 AU, with the same CZ at 3 AU as before. Figure~\ref{fig:influence_of_MMR} shows that in all cases the planets are trapped in MMRs between \MMR{11}{10} and \MMR{3}{2} (resonant order from 2 to 10).  As expected, the excited eccentricity decreases for higher order resonances and leads to a smaller inward shift of the planetary system.  The magnitude of the inward shift ranges from 0.2 to 1.5 AU.  In two simulations the planets started so close to each other that they were trapped in co-orbital (\MMR{1}{1}) resonance.  In these cases the eccentricities remained very low, and both planets migrated to the nominal (classical unshifted) CZ.

\section{Evolution with $>$2 planets}

We now turn our attention to the case of many planets in a disk.  We ran ten simulations each with two, three, five, and ten planets of $3 \mearth$ each.  The planets were placed randomly between 1 and 10 AU and were given low initial eccentricities and inclinations.  As before, each simulation was evolved for 3 Myr in a static disk.  

For the three-planet case we find three different outcomes.  First, the most likely outcome is that the three planets can be trapped in a resonant chain and migrate inward together to a zero net torque zone.  This zone is typically between 2 and 2.5 AU. 
The eccentricities of the three planets are not the same: the middle planet is generally more excited.  In the second most likely outcome, two planets are trapped in MMR and shifted inward, but a third, outer planet is too far away to be trapped in a resonant chain.  The outer planet remains at the nominal CZ at 3 AU, while the inner planets end up at a new shifted zero torque zone.  Finally, in some cases collisions occur between two planets and the system reverts to the two planet, unequal-mass case seen in \reffig{fig:mass_ratio_final_pos}.  

\begin{figure}[htb]
\centering
\includegraphics[width=\linewidth]{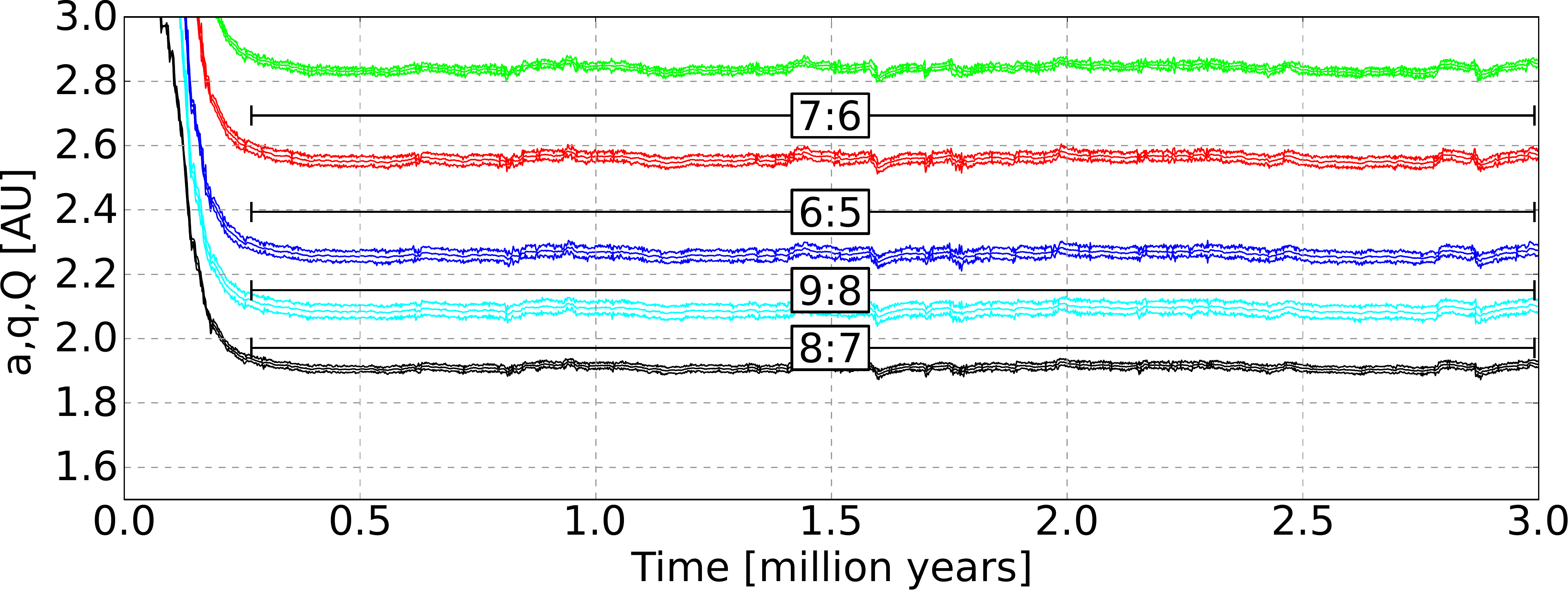}\\
\includegraphics[width=\linewidth]{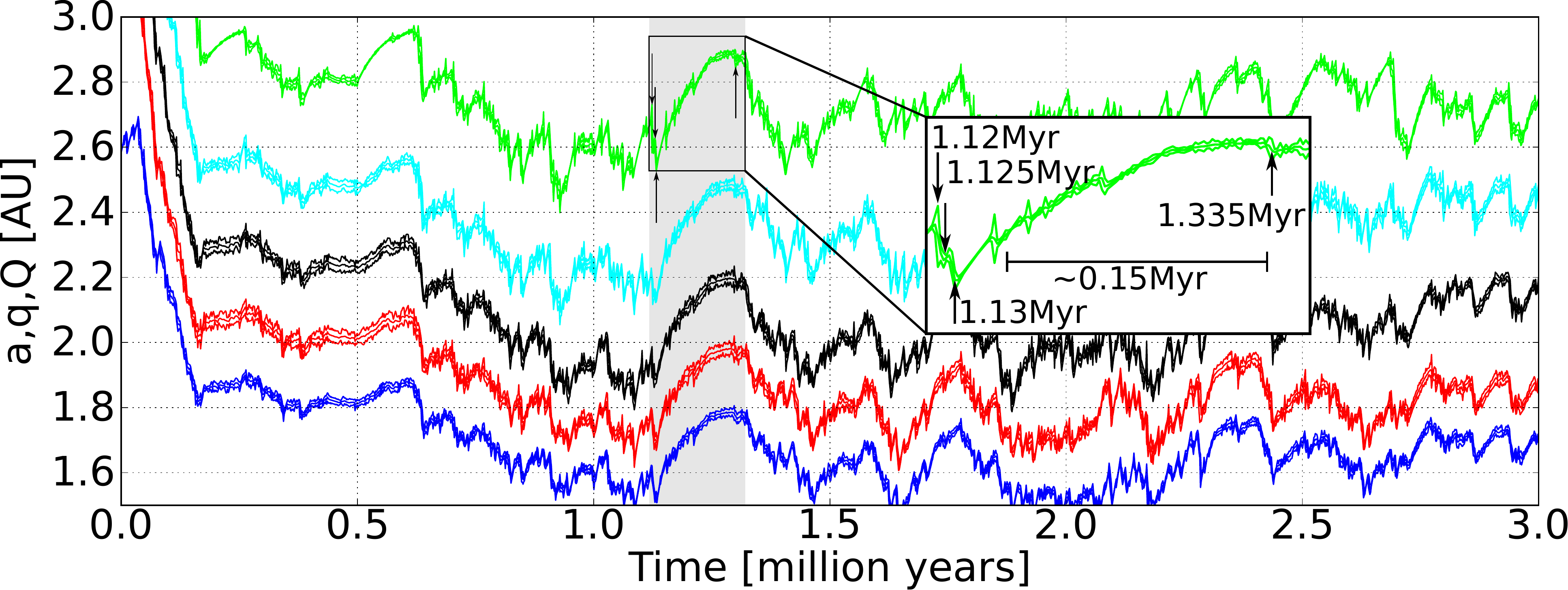}
\caption{Two example simulations with 5 planets, one that is quite stable even if short resonance breaking occurs (top panel)  and one that exhibits sustained chaotic behavior (bottom).}
\label{fig:timed-resonance-unstable}
\end{figure}

For the case of five or ten planets the situation is more complex.  The five-planet systems become trapped in chains of MMRs and migrate inward to an equilibrium position with zero net torque.  However, perturbations between planets add a stochastic element to the planets' behavior.  Even the most stable systems include periods during which the planets all drift radially in the same direction.  These periods are triggered by the breaking of a resonance between two planets that propagates as a perturbation through the whole system.  The amplitude and frequency of these chaotic periods vary from case to case.  For example, in the top simulation from \reffig{fig:timed-resonance-unstable} the resonant chain undergoes several small hiccups but the drift is small compared with the typical inter-planetary spacing.  In contrast, the perturbations in the simulation in the bottom panel of \reffig{fig:timed-resonance-unstable} are much stronger.  For instance, we consider the interval between roughly 1.1 and 1.3 Myr in the bottom case.  At 1.12 Myr the two outer planets become trapped in \MMR{4}{3} MMR. They migrate inward due to the resonant eccentricity increase.  This perturbation propagates to the inner system, and the eccentricity increase causes the planets to migrate inward all together. Five thousand years later, the outer planets leave and then re-enter the \MMR{4}{3} MMR, again perturbing the system. Finally, at 1.13 Myr the outer two planets leave the \MMR{4}{3} MMR for good.  As they leave the MMR, the two planets act like isolated bodies and migrate toward the nominal convergence zone with near zero eccentricities.  Without the net negative torque supplied by the outer two planets, the inner planets reactively migrate outward toward an effective three-planet-system zero torque zone.  This is the cause of the abrupt outward migration of all five planets.  However, the two outer planets quickly find themselves in the \MMR{5}{4} MMR.  Over the subsequent 0.15 Myr interval, they are periodically trapped in \MMR{5}{4} MMR but outward migration continues because most of the time they are not in MMR and their eccentricity is low.  The system-wide outward migration stops at 1.335 Myr when the outer planets cross over the \MMR{5}{4} MMR and are trapped in \MMR{6}{5} MMR. This configuration stabilize the system, excites the outer planets' eccentricities, and causes the whole system to migrate inward again, which signals the end of this particular meta perturbation. 

The rest of the evolution is composed of the same kinds of perturbations.  Perturbations come from planets entering or leaving resonances, then propagate between planets.  When leaving a resonance, planets' eccentricities usually drop, causing outward migration.  Conversely, being trapped in resonance excites a sustained eccentricity that leads to inward migration.  Via these perturbations, entire systems of planets undergo modest-scale chaotic migration due to the difficulty of maintaining resonant chains over long times.  Each of the five planet systems we simulated remained stable, in that no collisions occurred, but the amplitude of the chaotic migration varied from case to case; the two examples from \reffig{fig:timed-resonance-unstable} show the extremes. The ten-planet simulations were even more chaotic, and collisions did occur.  

The most important factor in determining the amplitude of chaotic oscillations is the order of the resonances. Low-order resonances maintain higher eccentricities and are less stable because they are sensitive to eccentricity variations. On the other hand, high-order resonances maintain lower eccentricities, and eccentricity perturbations are less important. For instance, in the system in the top panel of \reffig{fig:timed-resonance-unstable}, perturbations are quickly smoothed out, whereas in the system in the bottom panel the frequency of perturbations is high enough that the system does not tend to a stably evolving configuration.  Therefore, in this context a more compact system of planets is more stable than an extended system, which is the opposite of the pure gravitational situation \citep{marchal1982hill}.  

\section{Discussion}

We have shown that planets do not actually converge in convergence zones (CZs).  Instead, embryos rapidly migrate toward the CZ and are trapped in chains of MMRs.  This causes their eccentricities to increase and remain high enough to attenuate the corotation torque. The zero-torque equilibrium point of the resonant chain of planets is determined by the sum of each individual planet torque (sum of attenuated corotation torque and unattenuated differential Linbdlad torque). In practice, the effective zero net torque zone is shifted inward and is most strongly determined by the CZ of the most massive planet in the resonant chain. It is not a true CZ because each planet feels a different CZ (depending on its eccentricity).

The inward shift exists because planets' eccentricities are sustained by resonant perturbations.  The amplitude of a planet's eccentricity is a result of the competition between resonant excitation and eccentricity damping by the disk.  
For sufficiently high sustained eccentricities, an entire resonant chain of planets can migrate all the way in to the inner edge of the disk.  Changing the disk properties may thus change the typical sustained eccentricities by affecting the eccentricity damping timescale.  However, changing the disk properties affects other properties of the system such as the torque profile.  In changing the properties of the disk, it is not straightforward to assess the effect on the planets' evolution, as planets may be trapped in different resonances with different eccentricity excitations and even different stability criteria.  

The CZ depends on the disk parameters such as the viscosity, temperature and surface density profiles \citep[e.g.][]{paardekooper2011torque}. Here we have used a disk profile that is motivated by complex models but is nonetheless artificial.  Even if the results shown use a particular model, they are robust when we vary the torque shape in function of the orbital distance, as long as a CZ zone exists. In a more realistic disk we would expect a few differences.  First, there may be multiple CZs in the same disk owing to different effects \citep{lyra2010orbital, hasegawa2011origin}.  Second, mass-dependent convergence zones may exist in the outer part of the disk, where this mechanism should be less efficient because planets do not migrate to the same location in the disk and may not produce the resonant chains essential for our mechanism to occur (Cossou et al, in preparation).  Third, as the disk dissipates, so does the torque profile and the location of CZs \citep{lyra2010orbital, horn2012orbital}.  Finally, turbulence is thought to be common in real protoplanetary disks \citep{armitage2011dynamics}.  Although turbulence does not affect the long-term evolution of a single planet in a radiative disk \citep{pierens2012protoplanetary}, we expect it to modify resonant capture and eccentricity evolution \citep[see][]{pierens2011dynamics}.  

\begin{acknowledgements}
We thank B. Bitsch, F. Selsis, F. Hersant, and an anonymous referee for stimulating discussions.  We acknowledge funding from the CNRS's PNP program and the Conseil Regional d'Aquitaine. Computer time for this study was provided by the computing facilities MCIA (M\'{e}socentre de Calcul Intensif Aquitain) of the Universit\'{e} de Bordeaux and of the Universit\'{e} de Pau et des Pays de l'Adour.  
\end{acknowledgements}

\bibliographystyle{aa}
\bibliography{shifted}

\end{document}